\documentclass[fleqn]{article}
\usepackage{amsmath,amssymb}
\usepackage{epsfig,graphicx}

\catcode`@=11 \@addtoreset{equation}{section} \catcode`@=12

\begin{document}
\title{\vskip-1.7cm
\begin{flushright}
{\normalsize
CERN-PH-TH-2014-202}
\end{flushright}
\vspace{0.5cm}
\bf  Holography beyond conformal invariance and AdS isometry?\footnote{Prepared for a special issue of JETP dedicated
to the 60th birthday of Valery Rubakov.}}
\date{}
\author{A.O.Barvinsky}
\maketitle
{\em
Theory Department, Lebedev
Physics Institute, Leninsky Prospect 53, Moscow 119991, Russia}\\
\centerline{\em and}\\
\centerline{\em Theory Division, CERN, CH-1211 Geneva 23, Switzerland}\\

\begin{abstract}
We suggest that the principle of holographic duality can be extended beyond conformal invariance and AdS isometry. Such an extension is based on a special relation between functional determinants of the operators acting in the bulk and on its boundary, provided that the boundary operator represents the inverse propagators of the theory induced on the boundary by the Dirichlet boundary value problem from the bulk spacetime. This relation holds for operators of general spin-tensor structure on generic manifolds with boundaries irrespective of their background geometry and conformal invariance, and it apparently underlies numerous $O(N^0)$ tests of AdS/CFT correspondence, based on direct calculation of the bulk and boundary partition functions, Casimir energies and conformal anomalies. The generalized holographic duality is discussed within the concept of the ``double-trace" deformation of the boundary theory, which is responsible in the case of large $N$ CFT coupled to the tower of higher spin gauge fields for the renormalization group flow between infrared and ultraviolet fixed points. Potential extension of this method beyond one-loop order is also briefly discussed.
\end{abstract}

\maketitle

\section{Introduction}
\hspace{\parindent}
It is a great pleasure to write this paper dedicated to Valery Rubakov on the occasion of his sixties birtday. Our scientific careers have started simultaneously when we were students at the Moscow University and shared common interests in physics -- classical and quantum gravity -- and invariably pursued these interests, in our own ways and styles, throughout the years to come. In particular, the results of this work have been conceived in the course of discussions, when Valery suggested to work out the covariant method of calculating radiative corrections in brane gravity models \cite{Rubakov} as a means of establishing applicability limits of perturbation theory. By the time when this method has got ready for use the peak of interest in brane models was basically over, and interests of scientific community have shifted to other areas, not the least of those being the idea of holographic duality and AdS/CFT correspondence. Interestingly, that old method seems to find now application in this field, and, I hope, Valery would be amused to see how his suggestions get incarnation in this nonperturbative concept of high-energy physics.

The idea of holographic duality between the $d$-dimensional conformal field theory (CFT) and the theory in the $(d+1)$-dimensional anti-de Sitter (AdS) spacetime that initially began with supersymmetric models of $N\times N$-matrix valued fields \cite{AdS/CFT0,Witten,LiuTseytlin} was later formulated for much simpler ``vectorial" models without the need in supersymmetry \cite{vectorial}. These models have an infinite tower of nearly conserved higher spin currents and in this way naturally lead to the corresponding tower of massless higher spin gauge fields. Therefore holography concept implies that the dual theory should contain these fields in AdS spacetime, thus forming Vasiliev theory of non-linear higher spin gauge fields \cite{Vasiliev,Vasilievetal} which necessarily imply infinite set of those, because the principle of gauge invariance for spins $s>2$ cannot be realized for a finite tower of spins. In contrast to original supersymmetric models in which AdS/CFT correspondence was checked for supersymmetry protected correlators, in vectorial models holographic duality underwent verification by numerous nontrivial calculations which go beyond simple kinematical or group-theoretical reasoning and extend from the tree level $O(N^1)$ to the ``one-loop" order $O(N^0)$.

In particular, the calculation of the $U(N)$ singlet scalar CFT partition function on $S^1\times S^2$ was shown to agree with the corresponding higher spin partition function calculation in $AdS_4$ \cite{ShenkerYin}, this result being extended to the $O(N)$ singlet sector of scalar CFT \cite{collectivefield}. Then these results were confirmed and extended to arbitrary dimensions in \cite{GK-GKS} including comparison of thermal and Casimir energy parts of partition functions in $CFT_d$ and $AdS_{d+1}$ in \cite{GiombiKlebanovTseytlin}. Vanishing Casimir energy in odd-dimensional theory (associated with the absence of the conformal anomaly) implies the same on the AdS side, which is nontrivial because it implies infinite summation over the tower of higher spin gauge fields -- the property that was observed in $d=4$ on the $AdS_5$ side \cite{GKPST} and confirmed by explicit summation of conformal anomaly coefficients $a_s$ for conformal higher spin fields on the $S^4$ side \cite{Tseytlin}. The list of similar results agreeing on both sides of $AdS_{d+1}/CFT_d$ correspondence was extended in \cite{GiombiKlebanovTseytlin}.

A special class of holographic dualities is associated with the so-called double-trace deformations of the scalar CFT \cite{double-trace} which generates its renormalization group flow from the IR fixed point (free CFT) to the UV fixed point \cite{GubserKlebanov}. The associated holographic dual of this RG flow in AdS spacetime is the transition between two different boundary conditions on the dual massless gauge fields of higher spins at the AdS boundary \cite{GubserKlebanov,GKPST}.

Variety of these miraculous coincidences and gradually extending area of validity of duality relations (from supersymmetric models to non-supersymmetric ones, from lower spacetime dimensions and lower spins to higher ones, from divergent and Casimir energy parts of partition functions to their thermal parts, from bosons to fermions, etc.) imply that there should be some deep functional reasons underlying all this and perhaps even allowing one to extend holographic duality beyond AdS isometry and conformal invariance. The goal of this paper is to show that this is indeed possible. Within the class of holographic dualities associated with the double-trace deformation of CFT there exist universal relations for one-loop functional determinants of local and nonlocal operators on generic $(d+1)$-dimensional spacetime and its $d$-dimensional boundary \cite{qeastb} which guarantee this duality irrespective of the background geometry and conformal invariance. The only condition that relates $(d+1)$-dimensional and $d$-dimensional theories is that at the tree level the boundary theory is induced from the bulk by the Dirichlet boundary value problem -- then their one-loop quantum corrections dutifully match. The proof of this statement is based on linear algebra of (pseudo)differential operators and sequence of Gaussian functional integrations. When the theory has a small parameter $1/N$ playing the role of semiclassical Planck constant, this sequence of integrations might apparently be extended to holographic duality beyond one-loop order $O(N^0)$.

\section{Double-trace deformation of CFT and AdS/CFT correspondence}
\hspace{\parindent}
Double-trace deformation \cite{double-trace} of large $N$ CFT of scalar fields $\Phi^i(x)$, $i=1,...N$, by the square of the $O(N)$ invariant single-trace scalar operator $J(x)=\Phi^i(x)\Phi^i(x)$, $S_{CFT}(\Phi)\to S_{CFT}(\Phi)-\frac1{2f}\int dx\,J^2(x)$,
leads to the renormalization group flow between the IR fixed point of free CFT and its UV fixed point. In the limit of large $N$ it was clearly demonstrated by using the Habbard-Stratonovich transformation as follows \cite{GubserKlebanov}.

Let us consider the generating functional $Z_{CFT}(\varphi)$ of the correlators of $J$ for the perturbed theory with sources $\varphi$, $Z_{CFT}(\varphi)=\int d\Phi\,\exp\left(-S_{CFT}(\Phi)+\frac12\,J(\Phi) \mbox{\boldmath$f$}^{-1}J(\Phi)+\varphi J(\Phi)\right)$, so that
    \begin{eqnarray}
    &&\frac{Z_{CFT}(\varphi)}{Z_{CFT}(0)}
    =\left\langle\,\exp\left(\frac12\,{\hat J} \mbox{\boldmath$f$}^{-1}{\hat J}+\varphi\hat J\right)
    \right\rangle_{CFT}\equiv\left\langle\,
    e^{\,\varphi {\hat J}}\,
    \right\rangle_{CFT}^f,       \label{Z_CFT}\\
    &&\hat J\mbox{\boldmath$f$}^{-1}
    \hat J=\int dx\,dy\,\hat J(x) \mbox{\boldmath$f$}^{-1}(x,y)\hat J(y), \quad
    \varphi\hat J=\int dx\,\varphi(x)\,\hat J(x).
    \end{eqnarray}
For sake of generality of our formalism in what follows we write the operator $\mbox{\boldmath$f$}= \mbox{\boldmath$f$}(x,y)$ in a rather general form even though in CFT models it is ultralocal, $\mbox{\boldmath$f$}(x,y)=f\delta(x,y)$, and we also use condensed notation omitting the sign of integration over $d$-dimensional coordinates. Functional dependence in the $d$-dimensional space will be denoted by round brackets, like $S_{CFT}(\Phi)\equiv S_{CFT}(\Phi(x))$, and the operators acting in this space, like $\mbox{\boldmath$f$}$, will be boldfaced.

Representing the quadratic in $J$ part in the exponential of (\ref{Z_CFT}) as a Gaussian integral over the auxiliary field $\phi$ -- Habbard-Stratonovich transform -- we have
    \begin{eqnarray}
    &&\left\langle\,e^{\,\varphi {\hat J}}\,\right\rangle_{CFT}^f=
    ({\rm det}\,\mbox{\boldmath$f$})^{1/2}\int d\phi\, \left\langle\,\exp\left(-\frac12\,\phi\,
    \mbox{\boldmath$f$}\,\phi+(\phi+\varphi)\,\hat J\right)\right\rangle_{CFT},             \label{Z^f_CFT}
    \end{eqnarray}
where ${\rm det}\,\mbox{\boldmath$f$}$ denotes the {\em functional} determinant of the operator $\mbox{\boldmath$f$}(x,y)$ on the space of functions of $d$-dimensional coordinates.

Assuming as usual in large $N$ CFT the vanishing expectation value of $\hat J$, $\langle\,\hat J\,\rangle=0$, and smallness of higher order correlators $\langle\hat J\hat J...\hat J\rangle$ at $N\to\infty$,
    \begin{eqnarray}
    &&\left\langle\,e^{\varphi\,\hat J}\right\rangle_{CFT}\simeq
    \exp\left(\,\frac12\,\varphi\,\big\langle\,\hat J\,\hat J\,\big\rangle\, \varphi\right)\equiv
    \exp\left(-\frac12\,\varphi\,
    \mbox{\boldmath$F$}\varphi\right),\\
    &&\big\langle\,\hat J(x)\,
    \hat J(y)\,\big\rangle=
    -\mbox{\boldmath$F$}(x,y),                \label{JJ}
    \end{eqnarray}
where $-\mbox{\boldmath$F$}$ is the notation for the undeformed two-point correlator of $J$, we have from the new Gaussian integration in (\ref{Z^f_CFT})
    \begin{eqnarray}
    \left\langle\,e^{\,\varphi {\hat J}}\,\right\rangle_{CFT}^f=({\rm det}\,\mbox{\boldmath$f$})^{1/2}\left({\rm det}\,\mbox{\boldmath$F_f$}\right)^{-1/2}\,
    \exp\left(-\frac12\,\varphi\frac1{\mbox{\boldmath$F$}^{-1}
    +\mbox{\boldmath$f$}^{-1}}\varphi\right),                  \label{CFTside}
    \end{eqnarray}
where
    \begin{eqnarray}
    \mbox{\boldmath$F_f$}\equiv\mbox{\boldmath$F$}
    +\mbox{\boldmath$f$}.                           \label{Ff}
    \end{eqnarray}

Therefore the correlator $\langle\,\hat J\hat J\,\rangle_{CFT}^f$ in the double-trace deformed CFT interpolates between the UV and IR fixed points of the theory
    \begin{eqnarray}
    \langle\,\hat J\hat J\,\rangle_{CFT}^f=
    -\frac1{\mbox{\boldmath$F$}^{-1}
    +\mbox{\boldmath$f$}^{-1}}\;\to\;
    \left\{
    \begin{array}{ll}
    -\mbox{\boldmath$F$}+..., & \quad\mbox{\boldmath$f$}^{-1}\mbox{\boldmath$F$}\ll 1 \\
    &\\
    -\mbox{\boldmath$f$}+\mbox{\boldmath$f$}(\mbox{\boldmath$f$}^{-1}
    \mbox{\boldmath$F$})^{-1}+...,&
    \quad\mbox{\boldmath$f$}^{-1}\mbox{\boldmath$F$}\gg 1
    \end{array}
    \right.\,.
    \end{eqnarray}
For ultralocal $\mbox{\boldmath$f$}=f\delta(x,y)$ in the CFT with the single-trace scalar operator $\hat J$ of dimension $\Delta$, the correlator $\langle\,\hat J\hat J\,\rangle_{CFT}=-\mbox{\boldmath$F$}$ in the coordinate and momentum representations behaves as $-\mbox{\boldmath$F$}\sim 1/|x-y|^{2\Delta}\sim 1/k^{d-2\Delta}$. Thus, the above two limits indeed correspond respectively to the UV, $f^{-1}\mbox{\boldmath$F$}\sim -1/fk^{d-2\Delta}\ll 1$, and IR, $f^{-1}\mbox{\boldmath$F$}\sim -1/fk^{d-2\Delta}\gg 1$, fixed points. In the IR limit the correlator (modulo the contact term $\mbox{\boldmath$f$}=f\delta(x,y)$) is dominated by the second term $\mbox{\boldmath$f$}(\mbox{\boldmath$f$}^{-1}\mbox{\boldmath$F$})^{-1}\sim 1/|x-y|^{1(d-2\Delta)}$ in the long distance regime $|x-y|\gg |f|^{1/(d-2\Delta)}$ \cite{GubserKlebanov}. The renormalization group flow interpolates between two phases in which the operator $J(x)$ has different dimension, $\Delta=\Delta_+$ in IR and $d/2-\Delta=\Delta_-$ in UV.

This double-trace deformation picture takes place also in context of the dual description of higher spin conformal gauge fields \cite{GubserKlebanov,GKPST}. Since the $O(N)$ or $U(N)$ scalar or fermion CFT has a tower of nearly conserved higher spin currents $J_{\mu_1...\mu_s}(x)$, their gauging results in the corresponding tower of higher spin gauge fields $\varphi^{\mu_1...\mu_s}(x)$
    \begin{eqnarray}
    J=J_{\mu_1...\mu_s}(x)\sim
    \Phi^i(x)\,\partial_{\mu_1}...\partial_{\mu_s}\!\Phi^i(x), \quad
    \varphi=\varphi^{\mu_1...\mu_s}(x).
    \end{eqnarray}
This class of theories was conjectured to be dual to Vasiliev theories of higher spin gauge fields in AdS, a very incomplete list of references being contained in \cite{KlebanovPolyakov,SezginSundell_LeghPetkou,GiombiYin,
MaldacenaZhiboedov,DidenkoSkvortsovMei,GelfondVasiliev}. The description of these dualities can be summarized as follows.

In $AdS_{d+1}$ with the coordinates $X\equiv X^A=X^1,...X^{d+1}$ there exist totally symmetric transverse gauge fields $\varPhi=\varPhi^{A_1...A_s}(X)$ with the quadratic action
    \begin{eqnarray}
    S_{d+1}[\,\varPhi\,]=\int\limits_{AdS} d^{d+1}X\,{\cal L}\big(\varPhi(X),\nabla\varPhi(X)\big)
    \end{eqnarray}
that generates linearized equations for massless spin $s$ tensor fields. The covariant form of this quadratic action is known \cite{MHS}, but its concrete expression will not be needed in what follows. At the boundary of $AdS_{d+1}$ which is either $R^d$ or $S^d$ (or $S^1\times S^{d-1}$ in the thermal case) and parameterized by coordinates $x\equiv x^\mu=x^1,...x^d$ via the embedding functions $X=e(x)$ the boundary values of the tangential components of $\varPhi$
    \begin{eqnarray}
    \varPhi|\equiv \varPhi^{\mu_1...\mu_s}(e(x))=\varphi^{\mu_1...\mu_s}(x),
    \end{eqnarray}
represent the gauge fields of the $d$-dimensional CFT, coupled to its conserved higher spin currents. Then the $AdS_{d+1}/CFT_d$ conjecture means that the generating functional of the correlators of conserved currents of the {\em undefomed} CFT living on the boundary $\partial(AdS_{d+1})$ can be obtained from the path integral of the dual theory of gauge fields in the $AdS_{d+1}$ spacetime subject to Dirichlet boundary conditions at this boundary
    \begin{eqnarray}
    &&\Big\langle\,e^{\varphi\hat J}\,\Big\rangle_{CFT}=
    \frac{\int\limits_{\varPhi|\,=\,\varphi}
    D\varPhi\,e^{-S_{d+1}[\,\varPhi\,]}}
    {\int\limits_{\varPhi|\,=\,0}^{\vphantom{1}}
    D\varPhi\,e^{-S_{d+1}[\,\varPhi\,]}}\,.   \label{AdS/CFT}
    \end{eqnarray}
In what follows we will always denote the restriction of the bulk quantity to the boundary by the vertical bar.

Using this relation in the right hand side of (\ref{Z^f_CFT}) we get
    \begin{eqnarray}
    &&\left\langle\,e^{\,\varphi {\hat J}}\,\right\rangle_{CFT}^f=
    ({\rm det}\,\mbox{\boldmath$f$})^{1/2}
    \frac{\int d\phi\, e^{-\frac12\,\phi
    {\footnotesize\mbox{\boldmath$f$}}
    \phi}\int\limits_{\varPhi|=\phi+\varphi} D\varPhi\,e^{-S_{d+1}[\,\varPhi\,]}}
    {\int\limits_{\varPhi|=0}^{\vphantom{1}} D\varPhi\,e^{-S_{d+1}[\,\varPhi\,]}}\nonumber\\
    &&\qquad\qquad\qquad
    =({\rm det}\,\mbox{\boldmath$f$})^{1/2}\,\,
    \frac{\;\int\limits_{{\;\rm all}\;\varPhi} D\varPhi\,\exp\Big(-S_{d+1}[\,\varPhi\,]
    -\frac12\,\big(\varPhi|-\varphi\big)
    {\mbox{\boldmath$f$}}\big(\varPhi|-\varphi\big)\,\Big)}
    {\int\limits_{\varPhi|=0}^{\vphantom{1}}
    D\varPhi\,e^{-S_{d+1}[\,\varPhi\,]}}\,,    \label{AdS/CFT10}
    \end{eqnarray}
where the total action in the integrand of the functional integral contains both the bulk part and the boundary part located at $\partial(AdS_{d+1})=M_d$
    \begin{eqnarray}
    &&S[\,\varPhi\,]\equiv S_{d+1}[\,\varPhi\,]+\frac12\,\big(\varPhi|-\varphi\big)
    {\mbox{\boldmath$f$}}\big(\varPhi|-\varphi\big)\nonumber\\
    &&\qquad\qquad
    =\frac12\int\limits_{AdS} d^{d+1}X\,\varPhi(X)\!
    \stackrel{\leftrightarrow}{F}\!(\nabla)\,\varPhi(X)
    +\frac12\,\int\limits_{M_d}d^dx\,\big(\varPhi|(x)-\varphi(x)\big)
    {\mbox{\boldmath$f$}}
    \big(\varPhi|(x)-\varphi(x)\big),          \label{action0}
    \end{eqnarray}
and the integration in the denominator runs over the fields $\varPhi$ both in the bulk and on the boundary. This means that the boundary conditions on the saddle point configuration $\varPhi_f$ are affected by the boundary part of the action (that is by $\mbox{\boldmath$f$}$ which is a kernel of the quadratic boundary action in (\ref{action0})) -- that is why we label it by the subscript $f$.

The kernel of the bulk Lagrangian in given by the second order operator $F(\nabla)$, whose derivatives $\nabla\equiv\partial_X$ are integrated by parts in such a way that they form bilinear combinations of first order derivatives acting on two different fields (this is denoted by left-right arrow over $F(\nabla)$). Integration by parts gives nontrivial surface terms on the boundary. In particular, this operation results in the Wronskian relations for generic test functions $\varPhi_{1,\,2}(X)$ on any spacetime domain $M_{d+1}$ with the boundary $\partial M_{d+1}$
    \begin{eqnarray}
    &&\int\limits_{M_{d+1}} d^{\,d+1}X
    \left(\,\varPhi_1\!\stackrel{\rightarrow}{F}\!(\nabla)\varPhi_2-
    \varPhi_1\!\stackrel{\leftarrow}{F}\!(\nabla)\,\varPhi_2\right)=
    -\int\limits_{\partial M_{d+1}} d^{\,d}x
    \left(\,\varPhi_1\stackrel{\rightarrow}{W}\!
    \varPhi_2-
    \varPhi_1\stackrel{\leftarrow}{W}\!
    \varPhi_2\right),                          \label{Wronskianrel1}\\
    &&\int\limits_{M_{d+1}} d^{\,d+1}X\,\varPhi_1\!
    \stackrel{\leftrightarrow}{F}\!\varPhi_2=
    \int\limits_{M_{d+1}} d^{\,d+1}X\,
    \varPhi_1(\stackrel{\rightarrow}{F}\!\varPhi_2)
    +\int\limits_{\partial M_{d+1}} d^{\,d}x\,\varPhi_1\!
    \stackrel{\rightarrow}{W}\!
    \varPhi_2\,\Big|\,.                \label{Wronskian2}
    \end{eqnarray}
Arrows everywhere here indicate the direction of action of derivatives either on $\varPhi_1$ or $\varPhi_2$. These relations can be regarded as a definition of the first order {\em Wronskian operator} $W=W(\nabla)$ for $F(\nabla)$. In simple models on AdS background and its conformal boundary, parameterized by coordinates $X^A=y,x^\mu$ it basically reduces to the normal to the boundary derivative, $W(\nabla)\sim\partial_y$.

The saddle point approximation for the path integral in the numerator of (\ref{AdS/CFT10}) is dominated by the contribution of a stationary point of the total action (\ref{action0}). In view of (\ref{Wronskian2}) the requirement of its vanishing first order variation contains bulk and surface terms,
    \begin{eqnarray}
    \int\limits_{AdS_{d+1}} d^{\,d+1}X\,
    \delta\varPhi\,(\stackrel{\rightarrow}{F}\!\varPhi)
    +\int\limits_{M_d} d^{\,d}x\,\delta\varPhi\left(\big(\!
    \stackrel{\rightarrow}{W}\!+
    {\mbox{\boldmath$f$}}\big)\,\varPhi\,|
    -{\mbox{\boldmath$f$}}\varphi\right)=0\,,    \label{variation}
    \end{eqnarray}
which must independently vanish, because $\delta\varPhi(X)$ is nonvanishing both in the bulk and at the boundary since $\varPhi$ is being integrated over at all spacetime points. Thus we get equations of motion for a stationary configuration $\varPhi_f(X)$ in the bulk and the boundary condition at $M_d$,
    \begin{eqnarray}
    F(\nabla)\,\varPhi_f(X)=0,  \quad
    \big(\!\stackrel{\rightarrow}{W}(\nabla)
    +{\mbox{\boldmath$f$}}\big)\,\varPhi_f\big|=
    {\mbox{\boldmath$f$}}\varphi.
    \end{eqnarray}
The latter is the generalized Neumann (or Robin) boundary condition involving the normal to the boundary derivative of $\varPhi(X)$ contained in ${W}(\nabla)$. On the contrary, $\mbox{\boldmath$f$}$ is entirely $d$-dimensional operator, which is ultralocal in the CFT theory of double-trace deformations, but we will keep it in what follows more general (being a differential or even nonlocal pseudo-differential) operator in $M_d$.  The solution to this boundary value problem can be given in terms of the Green's function $G_{N_f}(X,Y)$ of $F(\nabla)$ subject to this (homogeneous) Neumann boundary condition
    \begin{eqnarray}
    F(\nabla)\,G_{N_f}(X,Y)=\delta(X,Y),\quad
    (\,\stackrel{\rightarrow}{W}
    +\mbox{\boldmath$f$}\,)\,
    G_{N_f}(X,Y)\,\Big|_{\,X\in
    \partial M_{d+1}}=0,                      \label{3.10}
    \end{eqnarray}
and it reads
    \begin{eqnarray}
    &&\varPhi_f(X)=\int_b dy\,G_{N_f}(X,y)\,\mbox{\boldmath$f$}\varphi(y)\equiv
    G_{N_f}|\,\mbox{\boldmath$f$}\varphi.               \label{3.12}
    \end{eqnarray}
Here $G_{N_f}(X,y)\equiv G_f(X,Y)\,|_{\,Y=e(y)}$ is the notation for the boundary-to-bulk propagator -- Green's function with its second argument put on the boundary via the embedding function $Y^A=e^A(y^\mu)$. Its subscript indicates that this Green's function is determined by the generalized Neumann boundary conditions with a particular function $\mbox{\boldmath$f$}$.

Then the stationary (on-shell) value of the action (\ref{action0}) equals
    \begin{eqnarray}
    &&S[\,\varPhi_f\,]=\frac12 \int_b dx\,dy\,\varphi(x)\mbox{\boldmath$f$}
    \Big(\mbox{\boldmath$f$}^{-1}(x,y)-G_{N_f}(x,y)\Big)
    \mbox{\boldmath$f$}\varphi(y)\equiv
    \frac12\,\varphi\big[\mbox{\boldmath$f$}
    -\mbox{\boldmath$f$}\!\left.\left. G_{N_f}\right|\right|\mbox{\boldmath$f$}\,\big]\,
    \varphi\, ,             \label{3.13}
    \end{eqnarray}
where, in its turn,
    \begin{eqnarray}
    G_{N_f}(x,y)\equiv G_{N_f}(X,Y)\,|_{\,X=e(x),\,Y=e(y)}\equiv G_{N_f}||
    \end{eqnarray}
is the notation for the boundary-to-boundary propagator -- the restriction of the both Green's function's arguments to the boundary being denoted for brevity by two vertical bars. And again using condensed notations on the boundary we omitted in (\ref{3.12})-(\ref{3.13}) the sign of integration over the {\em boundary coordinates}.\footnote{It is useful to apply this DeWitt condensed notation for integral operations on the boundary, because these operations have properties of formal matrix contraction and multiplication.} Thus finally we have
    \begin{eqnarray}
    &&\int D\varPhi\,\exp\Big(-S_{d+1}[\,\varPhi\,]
    -\frac12\,\big(\varPhi|-\varphi\big)
    {\mbox{\boldmath$f$}}\big(\varPhi|-\varphi\big)\,\Big)\nonumber\\
    &&\qquad\qquad\qquad
    =\left(\,{\rm Det}_{N_f}F\right)^{-1/2}
    \exp\Big(-\frac12\,\varphi\big[\mbox{\boldmath$f$}
    -\mbox{\boldmath$f$}\!\left.\left. G_{N_f}\right|\right|\mbox{\boldmath$f$}\,
    \big]\varphi\Big).                            \label{3.14}
    \end{eqnarray}
where ${\rm Det}_{N_f}F=({\rm Det}\,G_{N_f})^{-1}$ is the bulk functional determinant of $F$ on the space of functions subject to the generalized Neumann boundary conditions (\ref{3.10}). The denominator of (\ref{AdS/CFT10}) is given of course by the functional determinant with the Dirichlet boundary conditions corresponding to $\varPhi|=0$,
    \begin{eqnarray}
    {\int\limits_{\varPhi|=0}^{\vphantom{1}}
    DA\,e^{-S_{d+1}[\,\varPhi\,]}}=\big({\rm Det}_D F\big)^{-1/2}.
    \end{eqnarray}
Functional determinants of operators acting in the $(d+1)$-dimensional bulk here and in what follows will be denoted by ``${\rm Det}$" with the subscript indicating a type of boundary conditions for the class of functions on which the determinant is calculated (in contrast to ``${\rm det}$" for the operators acting on the boundary).

Substituting these results in (\ref{AdS/CFT10}) we get
    \begin{eqnarray}
    &&\Big\langle\,e^{\varphi \hat J}\,\Big\rangle_{CFT}^f=
    ({\rm det}\,\mbox{\boldmath$f$})^{1/2}
    \left(\frac{{\rm Det}_{N_f}F}{{\rm Det}_D F}\right)^{-1/2}
    \exp\Big(-\frac12\,\varphi\big[\,\mbox{\boldmath$f$}
    -\mbox{\boldmath$f$}\!\left.\left. G_{N_f}\right|\right|\mbox{\boldmath$f$}\,\big]\varphi\Big),
    \end{eqnarray}
so that comparison of the exponentials and preexponential factors here and in (\ref{CFTside}) yields the following tree-level and one-loop relations
    \begin{eqnarray}
    &&\left.\left. G_{N_f}\right|\right|
    =\mbox{\boldmath$F$}_f^{-1},      \label{tree-rel}\\
    &&{\rm Det}_{N_f} F={\rm det}\,
    \mbox{\boldmath$F$}_{\!f}\;
    {\rm Det}_D F.                   \label{1loop-rel}
    \end{eqnarray}

These relations are direct consequences of AdS/CFT correspondence (\ref{AdS/CFT}) in the lowest two orders of $1/N$ expansion, but the logic of this derivation can be reversed. If we start with these relations, then the holographic duality will be enforced in this approximation. As we will shortly show now, a simple exercise on linear algebra and Gaussian integration provides a proof that these relations are very general and hold for a generic second order differential operator $F(\nabla)$ acting on an arbitrary spin-tensor field for a generic manifold with a boundary. It induces by a special rule the (generically nonlocal pseudodifferential) operator $\mbox{\boldmath$F$}_f$ acting on the boundary, which can be regarded as an inverse (boundary-to-boundary) propagator of the surface theory induced from the bulk theory. No particular geometry of the bulk spacetime or its boundary is assumed in this construction. All this means that the holographic duality between $d$ and $(d+1)$-dimensional theories can be extended beyond AdS isometries and conformal invariance with the single assumption that the $d$-dimensional theory is induced from the bulk theory by integrating out its bulk degrees of freedom.

\section{Holographic duality and induced boundary theory}
\hspace{\parindent}
For the equations (\ref{tree-rel})-(\ref{1loop-rel}) to hold the boundary operator $\mbox{\boldmath$F$}_f$ should be related to the operator $F(\nabla)$ acting in the bulk and relevant boundary conditions $N_f$ on $\partial M_{d+1}$. To establish this we address the duality relation (\ref{AdS/CFT}) at the tree level. For a quadratic $(d+1)$-dimensional action of the form
    \begin{eqnarray}
    S_{d+1}[\,\varPhi\,]=
    \frac12\int\limits_{AdS} dX\,\varPhi(X)\!
    \stackrel{\leftrightarrow}{F}\!(\nabla)\,
    \varPhi(X)                                   \label{action2}
    \end{eqnarray}
the tree-level holographic duality (\ref{AdS/CFT}) implies that
    \begin{eqnarray}
    \Big\langle\,e^{\varphi \hat J}\,
    \Big\rangle_{CFT}=
    e^{-S_{d+1}[\,\varPhi_D(\varphi)\,]}/e^{-S_{d+1}[\,\varPhi_D(0)\,]},           \label{AdS/CFT1}
    \end{eqnarray}
where $\varPhi_D(\varphi)$ is a solution of the problem, $F(\nabla)\,\phi_D(X)=0$, $\phi_D\,|=\varphi(x)$, with the inhomogeneous Dirichlet boundary conditions. In view of relations (\ref{Wronskianrel1}) and (\ref{Wronskian2}) this solution and its on-shell value of the action can be represented in terms of the Dirichlet Green's function $G_D(X,Y)$,
    \begin{eqnarray}
    F(\nabla)\,G_D(X,Y)=\delta(X,Y),\quad
    G_D(X,Y)\,\Big|_{\,X=e(x)}=0.            \label{3.18}
    \end{eqnarray}
as
    \begin{eqnarray}
    &&\varPhi_D(X)=-\int_{M_d} dy\;G_D(X,Y)\!
    \stackrel{\leftarrow}{W}\Big|_{\,Y=e(y)}\varphi(y)\equiv
    -G_D\!
    \stackrel{\leftarrow}{W}\!|\;\varphi,         \label{3.19}\\
    &&\nonumber\\
    &&S[\,\varPhi_D\,]=
    \frac12\,\int_{M_d} dx\,dy\,\varphi(x) \Big[\,
    -\stackrel{\rightarrow}{W}\!G_{D}\!
    \stackrel{\leftarrow}{W}(x,y)\,\Big]\,
    \varphi(y)\equiv
    \frac12\,\varphi\Big[\,-
    \stackrel{\rightarrow}{W}\!G_{D}\!
    \stackrel{\leftarrow}{W}\!||\Big]\varphi\,.      \label{3.20}
    \end{eqnarray}
The expression $\stackrel{\rightarrow}{W}\!G_{D}\!
\stackrel{\leftarrow}{W}\!||$ implies that the kernel of the Dirichlet Green's function is being acted upon both arguments by the Wronskian operators with a subsequent restriction to the boundary,
    \begin{eqnarray}
    \stackrel{\rightarrow}{W}\!G_{D}\!
    \stackrel{\leftarrow}{W}\!||\,(x,y)\equiv\;
    \stackrel{\rightarrow}{W}\!(\nabla_X\!)\,G_{D}(X,Y)\!
    \stackrel{\leftarrow}{W}\!(\nabla_Y\!)
    \,\Big|_{\,X=e(x),\,Y=e(y)}.             \label{8}
    \end{eqnarray}
The result (\ref{3.20}) is exactly the tree-level boundary effective action obtained from the original action (\ref{1}) by integrating out the bulk fields subject to fixed boundary values $\varphi(x)$, $\mbox{\boldmath$S$} (\,\varphi\,)=S\left[\,\phi_D(\varphi)\,\right]$. Respectively, the kernel of the quadratic form of (\ref{3.20}) in $\varphi$ is the inverse propagator of boundary theory
    \begin{eqnarray}
    \mbox{\boldmath$F$}\equiv\frac{\delta^2\mbox{\boldmath$S$}}
    {\delta\varphi\,\delta\varphi}=
    -\stackrel{\rightarrow}{W}\!G_{D}\!
    \stackrel{\leftarrow}{W}\!||\,,           \label{3.21a}
    \end{eqnarray}
which is generically a nonlocal operator in the space of boundary coordinates $x$. Thus, the generating functional of correlation functions in the {\em undeformed} CFT reads
    \begin{eqnarray}
    \Big\langle\,e^{\varphi \hat J}\,
    \Big\rangle_{CFT}=\exp\left(-\frac12\,\varphi \mbox{\boldmath$F$}\varphi\right), \quad
    \big\langle\,\hat J\hat J\,\big\rangle_{CFT}=-\mbox{\boldmath$F$},
    \end{eqnarray}
the two-point correlator of the $\hat J$, cf. equation (\ref{JJ}), being induced from the $(d+1)$-dimensional bulk. In fact, this is a basic relation of the linearized tree-level AdS/CFT correspondence that was checked in numeerous models starting with \cite{Witten,LiuTseytlin}. This fixes the boundary operator $\mbox{\boldmath$F$}_f=\mbox{\boldmath$F$}+\mbox{\boldmath$f$}$
in the right hand sides of our basic relations (\ref{tree-rel})-(\ref{1loop-rel}) in terms of the bulk operator $F(\nabla)$. Let us now go over to the proof of these relations.

\section{Functional determinants relations}
\hspace{\parindent}
The idea of the derivation of relations (\ref{tree-rel}) and (\ref{1loop-rel}), that was first given in \cite{qeastb}, is based on a sequence of Gaussian functional integrations. Any action $S[\,\varPhi\,]$ quadratic in its field $\varPhi(X)$ can give rise to two Gaussian functional integrals. One of them is of the form
    \begin{eqnarray}
    Z=\int\limits_{\rm all} D\varPhi\,\exp{(-S[\,\varPhi\,])},  \label{ZN}
    \end{eqnarray}
where integration runs over all fields both in the bulk and on its boundary, and another,
    \begin{eqnarray}
    Z(\,\varphi\,)=
    \int\limits_{\varPhi|\,=\,\varphi}
    D\varPhi\,\exp{(-S[\,\varPhi\,])},   \label{ZD}
    \end{eqnarray}
implies integration with fixed values of $\varPhi$ at the boundary. Obviously these path integrals are related by the equation $Z=\int d\varphi\,Z(\,\varphi\,)$, so that independent calculation of its left and right hand sides yields certain tree-level and one-loop relations. As we will see, under an appropriate choice of $S[\,\varPhi\,]$ they turn out to be exactly the ones advocated above.

Consider the bulk-boundary action of the field $\varPhi(X)$ in the $(d+1)$-dimensional (bulk) spacetime $M_{d+1}$ and its boundary $M_d=\partial M_{d+1}$,
    \begin{eqnarray}
    &&S[\,\varPhi\,]=\frac12\int\limits_{M_{d+1}} dX\,\varPhi(X)\!
    \stackrel{\leftrightarrow}{F}\!(\nabla)\,\varPhi(X)
    +\int\limits_{M_d}
    dx \left(\frac12\,\varphi(x)\,
    \mbox{\boldmath$f$}(\partial)\,
    \varphi(x)+j(x)\,
    \varphi(x)\right),                       \label{1}\\
    &&\varPhi\;\big|\equiv
    \varPhi(X)\,\big|_{\,\partial M_{d+1}}=
    \varPhi(e(x))=\varphi(x).                 \label{2}
    \end{eqnarray}
We remind that the boundary embedding into the bulk in terms of $x=x^\mu$ is denoted by $X^A=e^A(x^\mu)$ and as above the vertical bar denotes the restriction of the bulk quantity to the boundary. The field $\Phi(X)$ and the second-order differential operator $F(\nabla)$ have absolutely generic spin-tensor structure, and there are no restrictions on geometry of the bulk $M_{d+1}$ and its boundary $M_d$. Similarly to (\ref{Wronskian2}), in the bulk part the derivatives of $F(\nabla)$ are integrated by parts in such a way that they form bilinear combinations of first order derivatives. The boundary part of the action contains as a kernel some local or nonlocal (pseudodifferential) operator $\mbox{\boldmath$f$}= \mbox{\boldmath$f$}(\partial)$, $\partial=\partial_x$ acting in the space of $x$. In contrast to the bulk part integration by parts on the boundary is irrelevant for our purposes, because $M_d$ is considered to be either closed compact or having trivial vanishing boundary conditions at its infinity. $j(x)$ plays the role of sources conjugated to $\varphi(x)$ and located on the boundary.

The calculation of (\ref{ZD}) repeats the derivation of Sect.2 -- the answer is given by
    \begin{eqnarray}
    Z=\left(\,{\rm Det}_{N_f}F\right)^{-1/2}
    \exp(-S[\,\varPhi_f\,]),                   \label{3.8}
    \end{eqnarray}
where $\varPhi_f$ is a stationary point of the action (\ref{1})
satisfying the following problem with the inhomogeneous generalized Neumann boundary conditions
    \begin{eqnarray}
    F(\nabla)\,\phi_f(X)=0,  \quad
    \left.(\,\stackrel{\rightarrow}{W}
    +\,\mbox{\boldmath$f$}\,)\,\phi_f\,\right|+j(x)=0,  \label{3.9}
    \end{eqnarray}
and ${\rm Det}_{N_f}F$ denotes the bulk ($(d+1)$-dimensional) functional determinant of $F(\nabla)$ on the space of functions subject to these (homogeneous) boundary conditions.

Similarly to (\ref{variation}) the problem (\ref{3.9}) naturally follows from the action (\ref{1}) and Wronskian relations for $F(\nabla)$, because the variation of the action is given by the sum of bulk and boundary terms, which separately should vanish since the action should be stationary also with respect to arbitrary variations of the boundary fields $\delta\varphi$. The Neumann Green's function of this problem (\ref{3.10}) gives the solution to (\ref{3.9}) which in condensed notations of Sect.2 (cf. Eq.(\ref{3.12})) reads $\varPhi_f(X)=-G_{N_f}|\,j$ and gives rise to the on-shell value of the action as a functional of the boundary source $j(x)$
    \begin{eqnarray}
    &&S[\,\phi_f\,]=-\frac12 \int_b dx\,dy\,
    j(x)\,G_{N_f}(x,y)\,j(y)\equiv
    -\frac12 \left.\left.
    j\,G_{N_f}\right|\right|\,j\, .             \label{3.13a}
    \end{eqnarray}
Here again $G_{N_f}(x,y)\equiv G_{N_f}(X,Y)\,|_{\,X=e(x),\,Y=e(y)}\equiv G_{N_f}||$ is the notation for the boundary-to-boundary propagator -- the restriction of the both Green's function's arguments to the boundary being denoted for brevity by two vertical bars. To simplify the formalism we
omitted in (\ref{3.13a}) the sign of integration over the {\em boundary coordinates}.\footnote{It is useful to apply this DeWitt condensed notation for integral operations on the brane, because these operations have properties of formal matrix contraction and multiplication.} Thus finally we have
    \begin{eqnarray}
    Z=\left(\,{\rm Det}_{N_f}\,F\right)^{-1/2}
    \exp\Big(\,\frac12 \left.\left.
    j\,G_{N_f}\right|\right|\,j\Big).          \label{3.14a}
    \end{eqnarray}

Alternatively one can calculate the same integral by splitting the integration procedure into two steps -- first integrating over bulk fields with fixed boundary values followed by the integration over the latter. This allows one to rewrite the same result in the form $Z=\int d\varphi\,Z(\,\varphi\,)$, where the inner integral (\ref{ZD})
    \begin{eqnarray}
    Z(\,\varphi\,)\equiv\int\limits_{\varPhi|\,=\,\varphi}
    D\varPhi\,\exp{(-S[\,\varPhi\,])}=
    \left(\,{\rm Det}_D\,F\right)^{-1/2}
    \exp(-S[\,\varPhi_{D}]).                          \label{Z1}
    \end{eqnarray}
is given by the contribution of the solution of the Dirichlet problem (\ref{3.19}) with the Dirichlet Green's function $G_D(X,Y)$, cf. Eq.(\ref{3.18}). The corresponding on-shell action  equals
    \begin{eqnarray}
    S[\,\varPhi_D\,]=
    \frac12\,\varphi\Big[\,-
    \stackrel{\rightarrow}{W}\!G_{D}\!
    \stackrel{\leftarrow}{W}\!||+\mbox{\boldmath$f$}\,\Big]\,
    \varphi
    +j\,\varphi\equiv
    \frac12\,\varphi\,\mbox{\boldmath$F$}_{\!f}\,
    \varphi+j\,\varphi.                          \label{3.20a}
    \end{eqnarray}
Quadratic in $\varphi$ part here coincides with the induced action (\ref{3.20}) of Sect.3 modulo the additional $\mbox{\boldmath$f$}$-term.

Substituting (\ref{Z1}) with (\ref{3.20a}) into $Z=\int d\varphi\,Z(\,\varphi\,)$ we again obtain the gaussian integral over $\varphi$ which is saturated by the saddle point $\varphi_0$ of the above boundary action (\ref{3.20a}), $\varphi_0=-\mbox{\boldmath$F$}_f^{-1} j$, and the final result reads
    \begin{eqnarray}
    &&Z=\big(\,{\rm Det}_D\,F\,\big)^{-1/2}
    \big(\,{\rm det}\,\mbox{\boldmath$F_f$}
    \,\big)^{-1/2}\,\exp\left(\,\frac12\,j\,
    \mbox{\boldmath$F_f$}^{-1}j\right),          \label{3.22}
    \end{eqnarray}
where we remind that ``det" denotes the functional determinants in the
$d$-dimensional boundary theory.

In view of arbitrariness of the boundary source $j$, comparison of the tree-level and one-loop (preexponential) parts with those of (\ref{3.14a}) immediately yields two relations
    \begin{eqnarray}
    &&G_{N_f}\,||=
    \mbox{\boldmath$F$}_f^{-1}\equiv
    \Big[\,-
    \stackrel{\rightarrow}{W}\!G_{D}\!
    \stackrel{\leftarrow}{W}\!||+
    \mbox{\boldmath$f$}\,\Big]^{-1},       \label{2.23}\\
    %&&\nonumber\\
    &&{\rm Det}_{N_f}\,F=
    {\rm det}\,\mbox{\boldmath$F$}_f\;
    {\rm Det}_{D}F.                   \label{3.24}
    \end{eqnarray}
This is exactly the relations (\ref{tree-rel})-(\ref{1loop-rel}) that underly the dual AdS description of the double-trace deformation of CFT models. The one-loop order equation (\ref{3.24}) here relates the functional determinants of the bulk operator on different functional spaces defined by Neumann and Dirichlet boundary conditions and intertwines them via the determinant of the boundary operator.\footnote{This might perhaps be a field-theoretic analogue of Vasiliev's determinants relation in the operator algebra of conformal currents \cite{GelfondVasiliev} based on different star products -- the counterpart to different functional spaces on a field theory side.}

When applied to a large $N$ CFT, these relations describe deformation of the boundary CFT which induces the renormalization group flow from the infrared ($f=\infty$) to the ultraviolet ($f=0$) fixed points of this theory and generates the corresponding increase of the central charge \cite{GubserMitra} (or the conformal anomaly $a$-coefficient in 4D case \cite{KomargodskiSchwimmer}). From (\ref{3.24}) the change of $f$-parameter is determined by the ratio
    \begin{eqnarray}
    &&\frac{{\rm Det}_{N_{f_1}}F}
    {{\rm Det}_{N_{f_2}}F}=
    \frac{{\rm det}\,\mbox{\boldmath$F$}_{f_1}}
    {{\rm det}\,\mbox{\boldmath$F$}_{f_2}}
    \equiv\frac{{\rm det}\,\big(\,\mbox{\boldmath$1$}
    +f_1^{-1}\,\mbox{\boldmath$F$}\,\big)}
    {{\rm det}\,\big(\,\mbox{\boldmath$1$}
    +f_2^{-1}\,\mbox{\boldmath$F$}\,\big)},     \label{3.25}
    \end{eqnarray}
where in the second equality we took into account that for an ultralocal kernel $\mbox{\boldmath$f$}=f\delta(x,y)$ its determinant ${\rm det}\,\mbox{\boldmath$f$}=1$ (say in dimensional regularization) does not give any contribution. This is the relation that was formulated in \cite{HartmanRastelli,DiazDorn} as the ratio of the bulk theory partition functions with different values of the $f$-coefficient in terms of the $\langle\,\hat J\hat J\,\rangle_{CFT}$ correlator of the unperturbed boundary CFT, $-\mbox{\boldmath$F$}=\stackrel{\rightarrow}{W}\!G_{D}\!
\stackrel{\leftarrow}{W}\!||$.\footnote{To compare (\ref{3.25}) with the formalism of \cite{HartmanRastelli} one should bear in mind that our $\mbox{\boldmath$f$}$ is the negative inverse of $f$ in \cite{HartmanRastelli}, and our $\mbox{\boldmath$F$}$ is the negative of the $\langle\,\hat J\hat J\,\rangle_{CFT}$ correlator denoted in \cite{HartmanRastelli} by $G$.}

While the right hand side of this equation was derived on the CFT side by using the Hubbard-Stratonovich transform \cite{GubserKlebanov}, the left hand side equality was proven in \cite{HartmanRastelli} by using the expression for the functional determinant of the Sturm-Lioville operator in terms of its basis functions \cite{BarvKam,KirstenMcKane} or by explicit use of the operator spectra on the AdS background. On the contrary, the power of our result (\ref{3.24}) is that it holds for generic bulk-boundary backgrounds for operators $F(\nabla)$ and $\mbox{\boldmath$f$}$ of the most general type and admits any type of the covariant regularization for UV divergences \cite{qeastb}.

\subsection{The case of gauge theories}
\hspace{\parindent}
Important remark is that the functional determinant duality relation  (\ref{3.24}) applies also to gauge theories, which is the case of major interest for us, because our goal is the holographic duality for towers of higher spin fields in the bulk and its boundary. A potential difficulty here might be the fact that in the bulk the totally symmetric spin-$s$ fields $\varPhi^{A_1...A_s}(X)$ have bulk indices ranging over $d+1$ values, while the boundary fields $\varphi^{\mu_1...\mu_s}(x)$ have only $d$-dimensional tensor components, so that bulk $F$ and boundary $\mbox{\boldmath$F$}$ operators have essentially different spin structure. This controversy is reconciled, however, by noting that spin $s>0$ theories are gauge invariant under the transformations of the form
    \begin{eqnarray}
    &&\varPhi\to\varPhi^\Xi=\varPhi+\Delta^\Xi\varPhi,\quad
    \varphi\to\varphi^\xi=\varphi
    +\Delta^\xi\varphi,
    \quad\Xi^{\|}|=\xi,         \label{varphi_transform}\\
    &&\Delta^{\varXi}\varPhi^{A_1...A_s}(X)
    =\nabla^{(A_1}\varXi^{A_2...A_s)}(X),\\
    &&\Delta^{\xi}\varphi^{\mu_1...\mu_s}(x)
    =D^{(\mu_1}\xi^{\mu_2...\mu_s)},
    \end{eqnarray}
generated by the spin $s-1$ field $\Xi(X)$ with the tangential components $\Xi^{\|}=\xi(x)$ ($D_\mu$ denotes the covariant derivative on the boundary). The balance of physical degrees of freedom in the bulk and on the boundary is then maintained by imposing gauge conditions fixing these transformations. Background covariant gauges of the form $H(\varPhi)=H^{A_1...A_{s-1}}(X)\sim
\nabla_B\varPhi^{BA_1...A_{s-1}}(X)=0$ fix them incompletely -- there remain residual gauge transformations which are the zero modes of the second order {\em bulk Faddeev-Popov operator} $Q=Q^{A_1...A_{s-1}}_{B_1...B_{s-1}}$ defined by
    \begin{eqnarray}
    \Delta^\Xi H(\varPhi)=Q\Xi.
    \end{eqnarray}
These modes are parameterized by the boundary values $\Xi|=\xi(x)$ which perform a gauge shift (\ref{varphi_transform}) of the boundary fields $\varphi$. Therefore, these residual gauge transformations can be gauged out by imposing on $\varphi$ the boundary gauge conditions of the form $h(\varphi)=h^{\mu_1...\mu_{s-1}}(x)
\sim D_\nu\varphi^{\nu\mu_1...\mu_{s-1}}(x)$. In their turn, they generate a nondegenerate {\em boundary Faddeev-Popov operator} $\mbox{\boldmath$Q$}=\mbox{\boldmath$Q$}^{\mu_1...\mu_{s-1}}_{\nu_1...\nu_{s-1}}$ defined by
    \begin{eqnarray}
    \Delta^{\xi}h(\varphi)=\mbox{\boldmath$Q$}\xi.
    \end{eqnarray}
Altogether this is equivalent to introducing under the path integral sign the Faddeev-Popov gauge breaking factor $\delta\left[\,H(\,\varPhi\,)\,\right]\,
\delta(h(\varphi))\,M_{\,H,h}[\,\varPhi\,]$ with\footnote{The ghost factor we use here involves a generic gauge, whereas the works on higher spin gauge fields on AdS background \cite{GaberdielGrumillerVassilevich_Saha_GuptaLal} usually use a particular (DeWitt background covariant) gauge defined by the generator of the gauge transformation. Moreover, in \cite{GaberdielGrumillerVassilevich_Saha_GuptaLal} the power of the Faddeev-Popov determinant in the ghost factor is different, because all the determinants are defined on functional spaces of symmetric tensor fields constrained by conditions of transversality and tracelessness.}
    \begin{eqnarray}
    \left(\,M_{\,H,h\,}[\,\varPhi\,]\,\right)^{-1}=
    \int D\Xi\;\delta
    \left[\,H(\,\varPhi^\Xi)\,\right]\,
    \delta(h(\,\varphi^\xi))\equiv\left(\,{\rm Det}_N Q\,\right)^{-1}\,.
    \end{eqnarray}

Again, using an obvious relation $\int D\Xi\,(...)=\int d\xi\,\int_{\Xi|=\xi} D\Xi\,(...)$ -- that the integral over the full algebra of gauge transformations decomposes into the integration over the algebra in the bulk with fixed transformations on the boundary and the subsequent integration over these boundary transformations, we have for the Faddeev-Popov gauge fixing factor
    \begin{eqnarray}
    &&\int D\Xi\;\delta\left[\,H(\,\varPhi^\Xi)\,\right]\,
    \delta\big(h(\,\varphi^\xi)\big)=
    \int d\xi \,\delta\!\left(\,
    \mbox{\boldmath$Q$}\xi\,\right)
    \!\!\!
    \int\limits_{\footnotesize
    \begin{array}{c}
    \Xi|\,=\xi
     \end{array}}
    \!\!\!\!D\Xi\;\delta\!\left[\;Q\Xi\,\right]\,
    \nonumber\\
    &&\nonumber\\
    &&\qquad\qquad=\left(\,{\rm det}\,
    \mbox{\boldmath$Q$}\,\right)^{-1}
    \!\!\!\int\limits_{\footnotesize
    \begin{array}{c}
    \Xi|\,=0\\
     \end{array}}
    \!\!\!\!\!\!D\Xi\;\delta\!\left[\;Q\Xi\,\right]
    =\left(\,{\rm det}\,
    \mbox{\boldmath$Q$}\,\right)^{-1}
    \left(\,{\rm Det}_D Q\,\right)^{-1}.  \label{4.13}
    \end{eqnarray}
which similarly to (\ref{3.24}) factorizes into the product of the bulk Dirichlet and boundary counterparts. We can use the t'Hooft trick to convert delta-function type gauges into the bulk and boundary gauge breaking terms
    \begin{eqnarray}
    &&\delta\left[\,H(\,\varPhi\,)\,\right]\,
    \delta(h(\varphi))\;\to\;\exp\left(-\frac12\int d^{d+1}X\,H^2(\,\varPhi(X)\,)-\frac12\int d^dx\, h^2(\varphi(x))\right).
    \end{eqnarray}
They contribute to operators $F$ and $\mbox{\boldmath$F$}_f$ their respective gauge breaking parts and make both of them nondegenerate. Then, ultimately in higher spin gauge teories the relation (\ref{3.24}) for the dual one-loop prefactors takes the form
    \begin{eqnarray}
    \frac{{\rm Det}_N Q}{\big(\,{\rm Det}_{N_f}F\,\big)^{1/2}}=
    \frac{{\rm det}\,\mbox{\boldmath$Q$}}{\big(\,{\rm det}\,\mbox{\boldmath$F$}_f\big)^{1/2}}\;
    \frac{\,{\rm Det}_D Q}{\big(\,{\rm Det}_D F\big)^{1/2}},
    \end{eqnarray}
and it can again be laid in the basis of holographic duality. Details of this bulk-boundary factorization including the Ward identities, which guarantee gauge independence of both boundary and bulk factors in the right hand side of this relation (of the choice of $h(\varphi(x))$ and $H(\varPhi(X))$ respectively), can be found in \cite{qeastbg}. The analysis of \cite{qeastbg} was done for spin two case, but it can easily be extended to all $s$.

\section{Conclusions and discussion}
\hspace{\parindent}
Thus, we have a strong evidence that the holography principle extends beyond conformal symmetry and AdS-isometry of underlying theories. In the class of AdS/CFT dualities associated with the double-trace deformation of CFT holography is dutifully enforced at the one-loop level wherever the holographic duality holds at the tree level in the form of the boundary theory induced from the bulk via the Dirichlet boundary value problem. This opens prospects for the further progress in holographic concept. First, arbitrariness of the background gives a firm ground for the tree-level duality beyond quadratic approximation for the action of bulk and boundary theories. Second, an obvious identity
    \begin{eqnarray}
    \int\limits_{\footnotesize\rm all}D\varPhi\,e^{-NS[\,\varPhi\,]}=\int d\varphi\,\int\limits_{\footnotesize
    \begin{array}{c}
    \varPhi|\,=\varphi\\
     \end{array}}D\varPhi\,e^{-NS[\,\varPhi\,]}
    \end{eqnarray}
applied to a nonlinear bulk-boundary action with $1/N\to 0$ playing the role of $\hbar$,
    \begin{eqnarray}
    S[\,\varPhi\,]=\int\limits_{M_{d+1}}d^{d+1}X\,
    \left(\frac12\,S_{(2)}\,\varPhi^2+\frac1{3!}S_{(3)}\varPhi^3+...\right)+
    \int\limits_{M_d}d^dx\,
    \left(\frac12\,\mbox{\boldmath$f$}_{(2)}\varphi^2+
    \frac1{3!}\,\mbox{\boldmath$f$}_{(3)}\varphi^3+...\right),
    \end{eqnarray}
suggests in higher-loop orders a sequence of new identities starting with (\ref{2.23})-(\ref{3.24}) and involving tree-level vertices of the action. This might help extending known results on AdS/CFT correspondence beyond one loop approximation.

Of course, there are certain limitations in applicability of the suggested method. It seems working only in one direction -- from a {\em local} theory in the bulk to a potentially nonlocal theory on the boundary (remember that the critical point of our derivation is a local second order in derivatives bulk operator $F(\nabla)$, the corresponding definition of its Wronskian operator $W(\nabla)$ and related Dirichlet and Neumann boundary value problems). At the same time, known numerous checks of AdS/CFT correspondence \cite{GiombiKlebanovTseytlin} start from the free local CFT at the boundary and match with partition functions of local, though apparently nonlinear, dual theories in the AdS bulk. In order to invert the setting in our holography derivation, perhaps one might start with the attempt of solving such a mathematical problem. Given a generic boundary action functional $\mbox{\boldmath$S$}(\varphi)$ of the field $\varphi(x)$ is it possible to find the functional of the bulk action $S[\,\varPhi\,]$ on $M_{d+1}$ whose on-shell value (subject to Dirichlet data on $\partial M_{d+1}$) would match with $\mbox{\boldmath$S$}(\varphi)$,
    \begin{eqnarray}
    \frac{\delta S[\,\varPhi_0\,]}{\delta\varPhi_0}=0,\quad
    \varPhi_0 \big|=\varphi\quad\to\quad S[\,\varPhi_0(\varphi)\,]=\mbox{\boldmath$S$}(\varphi).
    \end{eqnarray}
Apparently this problem does not have a unique solution, but the requirement of locality of $S[\,\varPhi\,]$ might restrict the class of possible solutions (if any), and then for a given boundary theory with the action $\mbox{\boldmath$S$}(\varphi)$ one may apply the above derivation by recovering first the local $S[\,\varPhi\,]$.

Practical importance of functional determinants relations (\ref{2.23})-(\ref{3.24}) is that they may be used in concrete physical problems. In \cite{qeastb} these relations were demonstrated to be useful for the derivation of surface terms of the Schwiger-DeWitt (Gilkey-Seely) coefficients in the heat kernel trace expansion -- the method important for the calculation of the Casimir energy, the boundary UV divergences, etc. The bulk-boundary/brane action (\ref{1}) finds application in the Randall-Sundrum brane world model \cite{RS} where the operator $\mbox{\boldmath$f$}$ is generated by the tension term on the brane. In the Dvali-Gabadadze-Porrati model \cite{DGP} $\mbox{\boldmath$f$}$ is a second order operator induced by the brane Einstein term, $\mbox{\boldmath$f$}(\partial)\sim\Box/\mu$ where $\mu$ is the DGP scale responsible for the cosmological acceleration \cite{Deffayet}. In context of Born-Infeld action in D-brane string theory with vector gauge fields $\mbox{\boldmath$f$}(\partial)$ is a first order operator \cite{open}.

Very interesting is the class of models in which holographic duality is associated not with the conformal infinity of the AdS spacetime, but rather gets realized for dynamically evolving (cosmological) branes which are nontrivially embedded into the spacetime with extra dimensions \cite{Rubakov,RS,DGP,Deffayet}. One such model is the large $N$ CFT driven 4D cosmology whose partition function serves as a source of quasi-thermal initial conditions for the Universe \cite{slih}. It is dual to the 5D Schwarzschild-de Sitter spacetime with the embedded spherical shell carrying the 4D Einstein action \cite{DGP/CFT} -- a realization of the dS/CFT correspondence \cite{dS/CFT} rather than AdS/CFT one. It is important that this 4D shell surrounding the Euclidean bulk black hole is not static, but rather its radius is periodically oscillating. This oscillatory dynamics in the bulk incorporates a dual description of the self-consistent 4D cosmological evolution driven by the large $N$ CFT in a quasi-thermal state, the amount of its quasi-equilibrium radiation being related to the bulk black hole mass. No doubt that there is a lot more new revelations and applications in store for us within this approach to perturbative and nonperturbative quantum gravity.

\section*{Acknowledgements}
The author strongly benefitted from fruitful discussions and correspondence with A.Tseytlin and M.Vasiliev.  This work was partly supported by the RFBR grant No. 14-02-01173.

\end{document}